\documentclass[10pt, ,letterpaper]{article}
\usepackage[top=0.85in,left=.8in,footskip=0.75in,marginparwidth=2in]{geometry}

\usepackage{xcolor}
\usepackage{soul}

\usepackage{cite}
\usepackage{amsmath}
\usepackage{lineno}

\usepackage{caption}
\usepackage{subcaption}

\usepackage{microtype}
\DisableLigatures[f]{encoding = *, family = * }

\usepackage{changepage}
\usepackage[aboveskip=1pt,labelfont=bf,labelsep=period,singlelinecheck=off]{caption}

\makeatletter
\renewcommand{\@biblabel}[1]{\quad#1.}
\makeatother

\usepackage{soul}
\usepackage{graphicx}
\usepackage{placeins}
\usepackage{mathptmx}
\usepackage{anyfontsize}
\usepackage{t1enc}
\usepackage{lastpage,fancyhdr,graphicx}
\usepackage{epstopdf}
\usepackage{cite} 
\usepackage{multirow}
\usepackage{url}
\usepackage{tabularx}
\usepackage{placeins}
\usepackage{multicol,lipsum}
\usepackage{rotating}
\usepackage{natbib}
\usepackage{float}
\usepackage{booktabs}
\usepackage{comment}

\fancyheadoffset[L]{2.25in}
\fancyfootoffset[L]{2.25in}

\usepackage{color}
\usepackage{float}

\definecolor{Gray}{gray}{.25}

\usepackage{graphicx}

\usepackage{sidecap}

\usepackage{wrapfig}
\usepackage[pscoord]{eso-pic}
\usepackage[fulladjust]{marginnote}
\reversemarginpar

\begin{document}
	\vspace*{0.35in}
	
	\begin{flushleft}
		{\Large
            \textbf\newline{Quantum-Enhanced Detection of Viral cDNA via Luminescence Resonance Energy Transfer Using Upconversion and Gold Nanoparticles}
		}
		\newline
		\\
            Shahriar Esmaeili\textsuperscript{1, 2},
		Navid Rajil\textsuperscript{1, 2},
     Ayla Hazrathosseini \textsuperscript{1, 2},
     Benjamin W. Neuman\textsuperscript{3},
		Masfer H. Alkahtani\textsuperscript{4},
  Dipankar Sen\textsuperscript{1, 2},
            Qiang Hu\textsuperscript{5},
            Hung-Jen Wu \textsuperscript{5},
            Zhenhuan Yi\textsuperscript{1, 2},
            Robert W. Brick \textsuperscript{1, 2},
            Alexei V. Sokolov\textsuperscript{1, 2, 6},
            Philip R. Hemmer \textsuperscript{1, 2, 6}, and
            Marlan O. Scully \textsuperscript{1, 2}

		\bigskip
	
	\textsuperscript{1}Institute for Quantum Science and Engineering, Texas A\&M University, College Station, TX 77843, USA\\
 \textsuperscript{2}Department of Physics and Astronomy, Texas A\&M University, College Station, TX 77843, USA\\
 \textsuperscript{3}Department of Biology, Texas A\&M University, College Station, TX 77843, USA\\
 \textsuperscript{4}King Abdulaziz City for Science and Technology (KACST), Riyadh 11442, Saudi Arabia\\
  \textsuperscript{5}Department of Chemical Engineering, Texas A\&M University, College Station, TX 77843, USA\\
\textsuperscript{6}Department of Electrical and Computer Engineering, Texas A\&M University, College Station, TX 77843, USA\\

		\bigskip

	\end{flushleft}
	\justify
	\section*{Abstract}

The COVID-19 pandemic has profoundly impacted global economies and healthcare systems, revealing critical vulnerabilities in both. In response, our study introduces a groundbreaking method for the detection of SARS-CoV-2 cDNA, leveraging Luminescence resonance energy transfer (LRET) between upconversion nanoparticles (UCNPs) and gold nanoparticles (AuNPs) to achieve an unprecedented detection limit of 242 femtomolar (fM). This innovative sensing platform utilizes UCNPs conjugated with one primer and AuNPs with another, targeting the 5' and 3' ends of the SARS-CoV-2 cDNA, respectively, enabling precise differentiation of mismatched DNA sequences and significantly enhancing detection specificity. Through rigorous experimental analysis, we established a quenching efficiency range from 10.4\% to 73.6\%, with an optimal midpoint of 42\%, thereby demonstrating the superior sensitivity of our method. By comparing the quenching efficiency of mismatched DNAs to the target DNA, we identified an optimal DNA:UCNP:AuNP ratio that ensures accurate detection. Our comparative analysis with existing SARS-CoV-2 detection methods revealed that our approach not only provides a lower detection limit but also offers higher specificity and potential for rapid, on-site testing. This study demonstrates the superior sensitivity and specificity of using UCNPs and AuNPs for SARS-CoV-2 cDNA detection, offering a significant advancement in rapid, accessible diagnostic technologies. Our method, characterized by its low detection limit and high precision, represents a critical step forward in managing current and future viral outbreaks, contributing to the enhancement of global healthcare responsiveness and infectious disease control.
\newline
\newline
	\textbf{Keywords:} Luminescence resonance energy transfer (LRET), Upconversion nanoparticles, SARS-CoV-2 cDNA, Quantum Sensing

\section{Introduction}\label{sec1}
The COVID-19 pandemic has brought unprecedented challenges to global health and the economy. Since the first case was identified in December 2019, the number of COVID-19 cases has rapidly increased worldwide, with a total of over 135 million confirmed cases and 2.9 million deaths reported as of April 2021 \citep{WHO2023}. Testing is crucial for patient management and pandemic control, including identifying and isolating infected individuals, tracing contacts, and monitoring the spread of the virus. However, the current state of SARS-CoV-2 detection is still fraught with challenges \citep{rajil2022quantum, rajil2021fiber}.

RT-PCR is considered the gold standard for SARS-CoV-2 testing due to its high sensitivity and specificity \citep{woloshin2020false}. However, it has some limitations, including the need for expensive equipment and skilled personnel, the length of the testing process, and a significant number of false-negative results \citep{he2020temporal, lan2020positive, xiao2020false}.

Nanotechnology-based biosensors, such as lateral flow assays, surface-enhanced Raman scattering, Luminescence resonance energy transfer (LRET), and electrochemical biosensors, present rapid, cost-effective, and user-friendly alternatives to traditional methods for detecting SARS-CoV-2 infections. For example, \cite{song2022multiplexed} developed a LRET-based biosensor for the rapid and precise detection of SARS-CoV-2 RNA through multiplexed analysis, achieving detection limits of 15 pM and 914 pM for ORF and N genes, respectively. Their method improves detection efficiency and reduces false negatives by analyzing two gene fragments concurrently. These advanced techniques enable researchers to overcome the limitations of existing diagnostic methods and enhance COVID-19 detection \citep{qiu2020dual, gupta2020nanotechnology, seo2020rapid, fabiani2021magnetic, moitra2020selective}. 

Luminescence resonance energy transfer (LRET), offering molecular scale information, is a powerful and versatile technique extensively used in biomedical and clinical applications. LRET relies on the non-radiative transfer of energy between a donor fluorophore and an acceptor chromophore, which typically occurs over distances in the 1-10 nm range. 
Its sensitivity to distance makes LRET an excellent tool for studying molecular interactions, conformational changes, and proximity-based assays \citep{clegg199218, roy2008practical, selvin2008single, chen2008versatile, hemmer2023opportunitie}. 
One of the applications of LRET in biomedical research is to monitor the presence or absence of specific nucleic acid sequences. The nucleic acid targets could cross-link donor and receptor particles upon nucleic acid hybridization, leading to LRET signals \citep{tyagi1996molecular, marras2006selection, tsourkas2003hybridization,sperling2010surface, hemmer2023engineering}.
Oligonucleotide probes could be covalently linked to donor and acceptor dyes via EDC sulfo-NHS activation chemistry. LRET-based DNA detection assays have been applied in various areas, including gene expression analysis, single nucleotide polymorphism (SNP) detection, and pathogen identification \citep{hermanson2013bioconjugate, didenko2001dna}.

Our study introduces a novel LRET-based method that employs nanotechnology-based biosensors to identify SARS-CoV-2 accurately. This innovative method utilizes upconversion and gold nanoparticles, a choice inspired by ultrasensitive detection methods previously reported by \cite{tsang2016ultrasensitive}. To further enhance the performance of our biosensors, we have adopted the 2,2'-[ethylenebis(oxy)] bisacetic acid (EBAA) method for synthesizing hydrophilic upconversion nanoparticles (UCNPs). This synthesis approach improves the stability and biocompatibility of the UCNPs and significantly boosts the LRET-based assays' overall efficacy.

Upconversion nanoparticles (UCNPs) are a class of nanomaterials that exhibit unique quantum properties, making them particularly suitable for advanced biosensing applications. These nanoparticles are capable of converting low-energy infrared photons into higher-energy visible or ultraviolet photons through a non-linear optical process known as upconversion. This process relies on the sequential absorption of multiple photons, which is inherently quantum mechanical. The resulting emission can be precisely tuned by altering the composition and structure of the UCNPs, allowing for highly sensitive and specific detection of target molecules \citep{su2017resonance}. In our study, the utilization of UCNPs enhances the luminescence resonance energy transfer (LRET) mechanism, enabling the detection of viral cDNA with remarkable precision and sensitivity.

In enhancing the design and functionality of our biosensors, we leveraged the unique optical properties of gold nanoparticles combined with the minimal autofluorescence and enhanced emission capabilities of upconversion nanoparticles. This strategic integration results in a detection system that exhibits exceptional sensitivity and selectivity for SARS-CoV-2 cDNA, capable of achieving a lower detection limit of 242 fM. Our assays have demonstrated a robust dose-response relationship, with a notable midpoint of quenching efficiency at 42\%, correlating to a DNA concentration of 36.54 pM. This precision in detection underscores the potential of our method to significantly outperform existing, potential future directions for developing and applying our LRET-based approach, the basic principles underpinning our approach, highlighting its potential advantages over other techniques, and sharing our experimental results in the LRET-based assessment of SARS-CoV-2 cDNA.

\section{Materials and Methods}
\subsection{Activation of Carboxyl-functionalized UCNPs with EDC/sulfo-NHS}
The synthesis of carboxyl-functionalized UCNPs was reported in the supplementary document. To activate COOH-UCNPs using EDC/NHS,
0.5 mg/ml of UCNPs was mixed with aqueous EDC/NHS solution \citep{dumelin2006selection, chen2008versatile}. Our 
COOH-UCNPs (LiYF$_4$(18\%), Er(1.5\%), Tm(0.5\%)@COOH), synthesized using the EBAA method with an average 11 nm in diameter, are detailed in the supplementary information. 2ml of COOH-UCNPs was mixed with 10ul of $0.3\ mg/\mu l$ NHS and 10ul of $0.2\ mg/\mu l$ M of EDC solutions. The solution is briefly vortexed and then vigorously shaken at 500 rpm for 0.5 hours using an Eppendorf MixMate or shaker. The activated particles are subsequently centrifuged at 5080 rcf (9000 rpm) for 10 minutes. Nearly 95\% of the supernatant is removed and replaced with fresh DI water. The particles are then resuspended by sonication for about 20 minutes and assessed for their dispersity using a 980 nm laser by holding the vial in front of the laser beam and visually inspecting the solution. More details of this part of the experiment are reported in the supplementary document.

\subsubsection{Conjugation of Activated-UCNPs with Amino-Modified Oligonucleotid}
The Carboxyl-functionalized UCNP (LiYF$_4$(18\%), Er(1.5\%), Tm(0.5\%)) is covalently conjugated with amino modified oligonucleotide probe. The ratio of UCNP to amino-modified oligonucleotide used in the study is 1:10. The molarity of the amino-modified oligonucleotide solution 
acquired from Integrated DNA Technologies (IDT) is 
100 $\mu M$.
The conjugation was performed by mixing 78 $\mu$l of the amino-modified oligonucleotide solution (equivalent to $4.67 \times 10^{13}$ oligonucleotides) with 1.3 mg of activated UCNP (equivalent to $4.67 \times 10^{12}$ individual particles). The solution mixture was then incubated overnight at 4°C with constant agitation. Subsequently, the particles were washed with DI water three times using centrifugation at 5080 rcf (9000 rpm) for 10 mins. More details of this part of the experiment are reported in the supplementary document.

\subsection{Conjugation of Thiol-Modified primer with AuNPs}

This protocol is used to label the 5nm Gold Nanoparticles (AuNP) with thiolated oligonucleotide. The ratio of AuNP to thiol-modified oligonucleotide used in the study is 1:10. The mass concentration of 5 nm AuNP is around 0.06 mg/ml, which we calculate the particle concentration as $4.43 \times 10^{13}$ particles/mL. The molarity of the thiol-modified oligonucleotide stock solution purchased from IDT 100 $\mu M$. 16 $\mu l$ of our thiol-modified oligonucleotide solution contains a total of $9.6 \times 10^{14}$ individual oligonucleotides. To conjugate the thiolated oligonucleotide with AuNP, we first centrifuged 30 $\mu l$ of TCEP at 50 g, removed the supernatant, and washed the gel twice with DDI water.
Subsequently, we added 16 $\mu l$ of thiolated oligo stock solution into the washed TCEP gel, followed by vortexing for 3 minutes and incubating it for 1 hour. 

The microtube was then centrifuged, and we recovered the supernatant containing reduced oligonucleotide. Then we added 2.17 ml of AuNP into the mixture of oligo and TCEP gel we recovered before and then incubated for 2 hours \citep{thermofisher2022}. The resultant mixture was washed three times using an ultracentrifuge (rotor TLA-110) at 110K RPM for 10 minutes to remove any excess oligos. Finally, the final pellet was dispersed in DI water for further analysis \cite{esmaeili2021monitoring}. More details of this part of the experiment are reported in the supplementary document.

\subsection{Detection of DNA}

The sensor system for this study was comprised of UCNPs-primer 2 and  AuNPs-primer 1 conjugates, as well as the target DNAs and the 24-base mismatched DNA (DNA-mmP1P2) and 12-base mismatch DNA (DNA-mmP1 and DNA-mmP2), 
The detection was conducted by mixing UCNP, AuNP, and DNA in DI water. The experiment consisted of two parts. First, we assessed the dose-response of the sensor by keeping the sensor's concentration constant while the final concentration of target DNA was reduced from $5.06~\mu M$ down to $5.06~fM$ (table \ref{table1}). Second, to test the specificity of the sensor in detecting SARS-CoV-2 cDNA, we tested 3 different mismatch bases, DNA-mmP1P2, DNA-mmP1, and DNA-mmP2, at the $5.06~\mu M$ concentration.

To commence the experiment, a combination of the target DNAs and amino-modified UCNPs were mixed and incubated at room temperature for 2 hours. After the incubation period, the solution was combined with a thiol-modified AuNP solution at a 1:5 number ratio and mixed thoroughly for 15 mins. The system was analyzed and measured for luminescence spectra during continuous-wave stimulation at 980 nm. Table \ref{table1} presents the quantities of the materials utilized in this experiment, including UCNPs, UCNP-conjugated primers, target DNA sequences, AuNPs, and AuNP-conjugated oligonucleotides.

\begin{table}[H]
\caption{Quantities of experimental materials used in the LRET-based SARS-CoV-2 cDNA detection study. The final concentration of DNA is varied while the number of upconversion nanoparticles (UCNPs), UCNP-conjugated primers, gold nanoparticles (AuNPs), and AuNP-conjugated primers remain constant at $4.7 \times 10^{12}$, $4.7 \times 10^{13}$, $2.35 \times 10^{13}$, and $2.35 \times 10^{14}$ respectively.}
\footnotesize
    \centering
\begin{tabular}{|l|l|}
 \hline
\textbf{Final Concentration of DNA [M]} & \textbf{Number of Target DNA} \\\hline
Cntl: 0                                & 0                             \\\hline
5.06 $\times 10^{-6}$                  & $3.6 \times 10^{14}$          \\\hline
5.06 $\times 10^{-8}$                  & $3.6 \times 10^{12}$          \\\hline
5.06 $\times 10^{-11}$                 & $3.6 \times 10^{9}$           \\\hline
5.06 $\times 10^{-12}$                 & $3.6 \times 10^{8}$           \\\hline
5.06 $\times 10^{-14}$                 & $3.6 \times 10^{6}$           \\\hline
5.06 $\times 10^{-15}$                 & $3.6 \times 10^{5}$           \\\hline
\end{tabular}
\label{table1}
\end{table}

\section{Results and Discussion}

\begin{figure}[H]
    \centering
    \begin{subfigure}[b]{0.9\textwidth}
        \includegraphics[width=\textwidth]{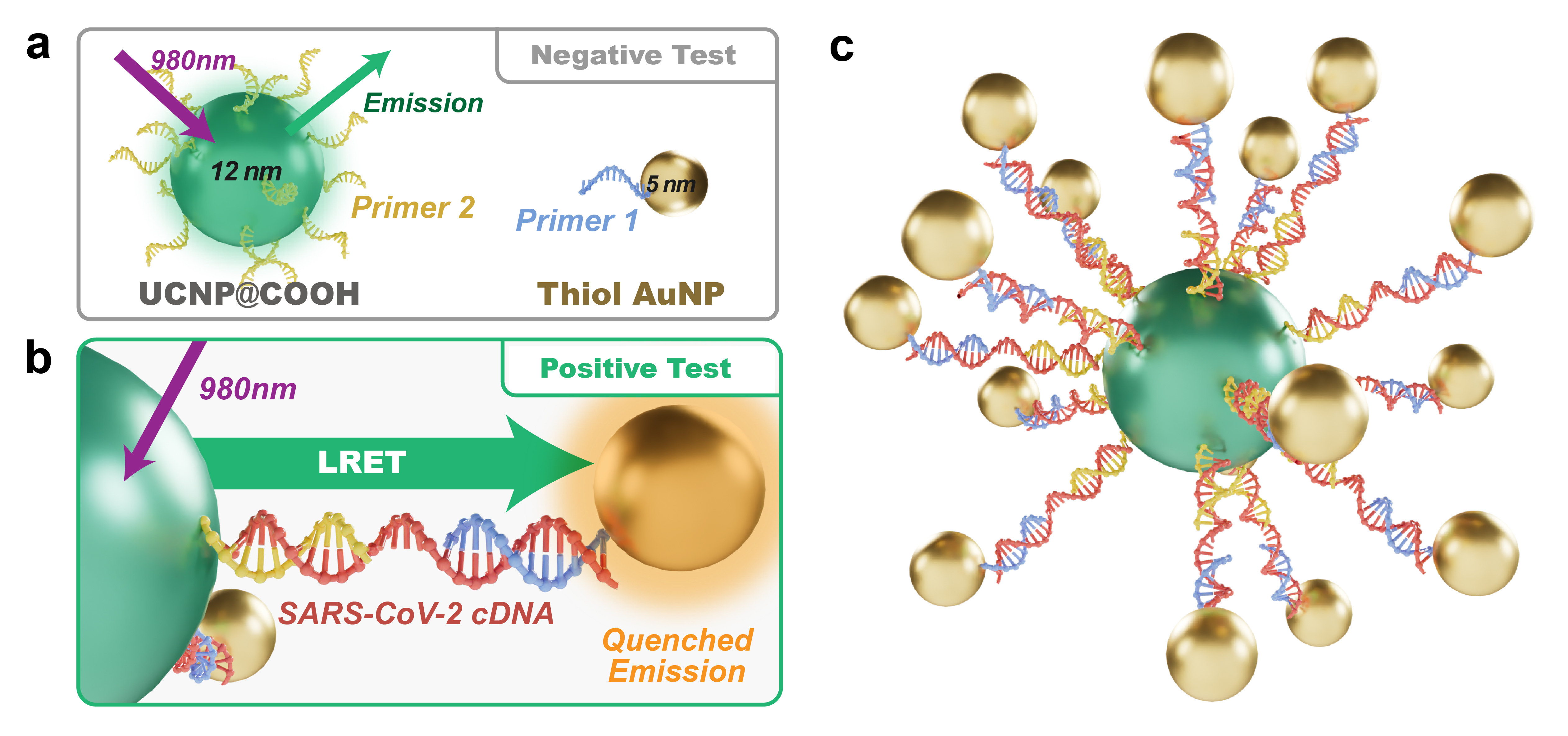}
        \label{fig3-A}
    \end{subfigure}
    \hfill
    \begin{subfigure}[b]{1\textwidth}
        \includegraphics[width=\textwidth]{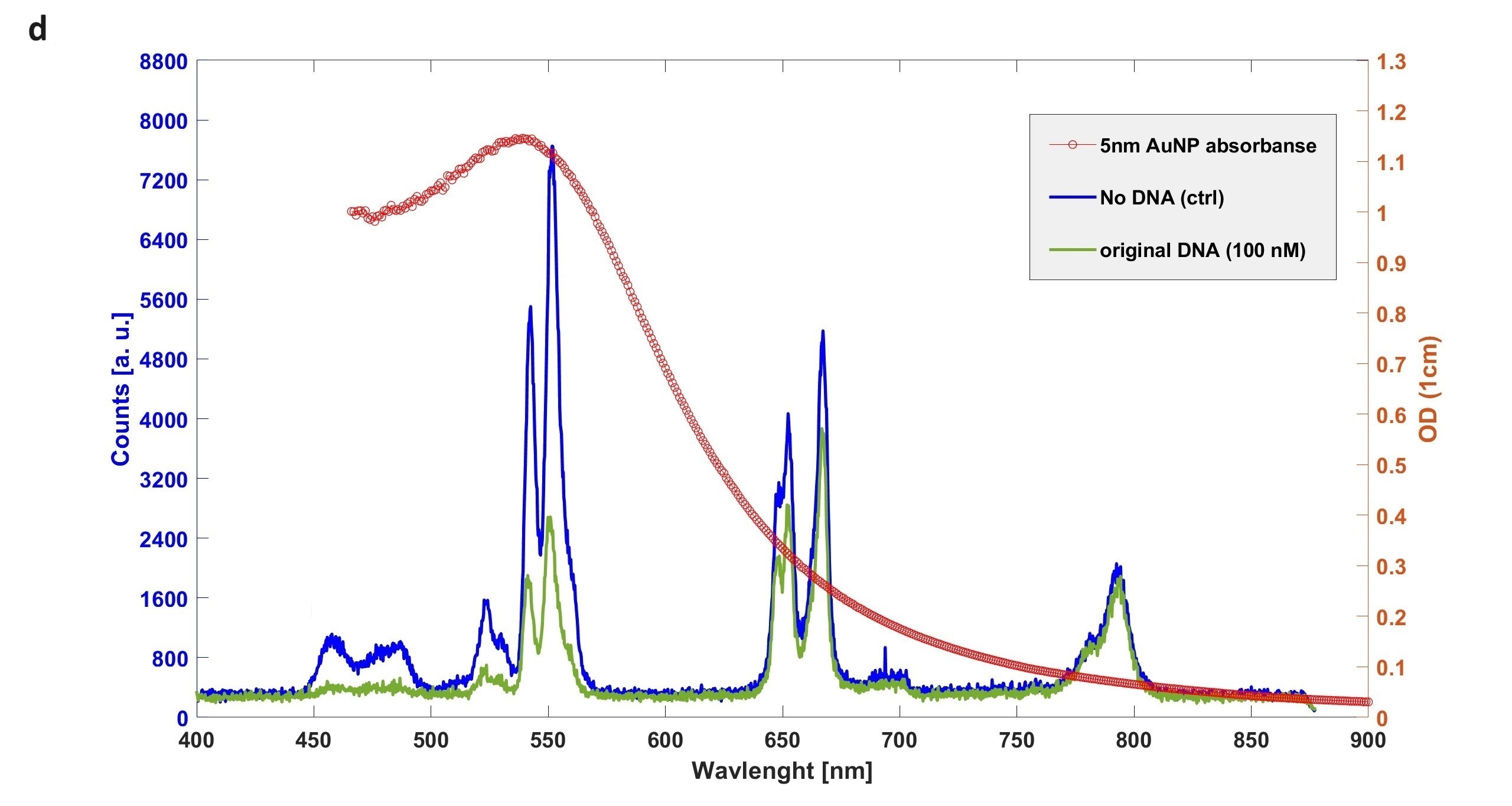}
        \label{fig3-B}
    \end{subfigure}
    \caption{Feasibility testing of the assay: The negative (control) test. (a) shows the schematic of the assay, where the DNA is absent, and the UCNP and AuNP will not bind, resulting in higher fluorescence intensity from UCNPs. (b) shows the schematic of the assay, where the DNA is present, and the UCNP and AuNP will bind, resulting in lower fluorescence intensity from UCNPs due to quenching caused by LRET coupling. (c) shows the 3D schematic of conjugated UCNP surrounded by conjugated AuNPs linked by SARS-CoV-2 cDNA. (d) shows an overlay of the AuNP absorption spectrum and UCNP fluorescence spectrum of the negative (control, no DNA) and positive tests (original DNA, 100 nM). The positive test shows lower fluorescence, particularly around the 550 nm wavelength, while the quenching effect is less pronounced at the 667 nm and 792 nm wavelengths.}
     \label{fig3}
\end{figure}

Before beginning the dose-response measurement, we examined the feasibility of this assay by testing both positive and negative (control) samples to ensure that the particles and the target \textit{SARS-CoV-2 cDNA} were binding correctly. The negative (control) test consisted of UCNPs conjugated with primer 2 and AuNPs conjugated with primer 1, to which we added $6~\mu l$ of DI water (Fig. \ref{fig3}).

For the positive test, we used the same concentration and volume of UCNPs conjugated with primer 2 and AuNPs conjugated with primer 1, but added $6~\mu l$ of \textit{SARS-CoV-2 cDNA} at a stock solution concentration of $10^{-4}\ M$. This resulted in a final DNA concentration of $5.06~\mu M$ (see Table \ref{table1} and Fig. \ref{fig3}).

In the absence of DNA, the AuNPs would not bind to the UCNPs through DNA-Primer hybridization. In this case, we expected higher fluorescent intensity from UCNPs (Fig. \ref{fig3}). In contrast, the proper sequence of DNA can cross-link UCNP and gold particles, leading to LRET coupling. Therefore, the quenching of fluorescent signals from the UCNPs would be observed (Fig. \ref{fig3}). The positive test clearly showed a lower fluorescent intensity, specifically at around 550 nm wavelength.

The major peaks we considered in this experiment are 457nm, 523nm, 550nm, 667, and 792nm. This selection allows us to have different parts of the absorption spectrum of AuNPs involved in the LRET coupling. The absorption cross-section of AuNPs is plotted with the spectrum of UCNPs in Fig. \ref{fig3} (d). Keeping in mind the absorption cross-section of AuNPs, it is noteworthy that the quenching in the 667 nm wavelength was less than that of 550 nm, and the quenching in the 792 nm wavelength was less than that of 650 nm \citep{esmaeili2022detection}. The reason for the difference in the quenching of 550, 667, and 792 nm wavelength can be explained by the gold nano particle's absorption spectrum. One can see that the maximum absorption for AuNPs of 5 nm is around 520 nm, while the absorption of 667nm and 792nm are significantly lower.
%

\begin{figure}[H]
		\centerline{\includegraphics[width=.9\textwidth,clip=]{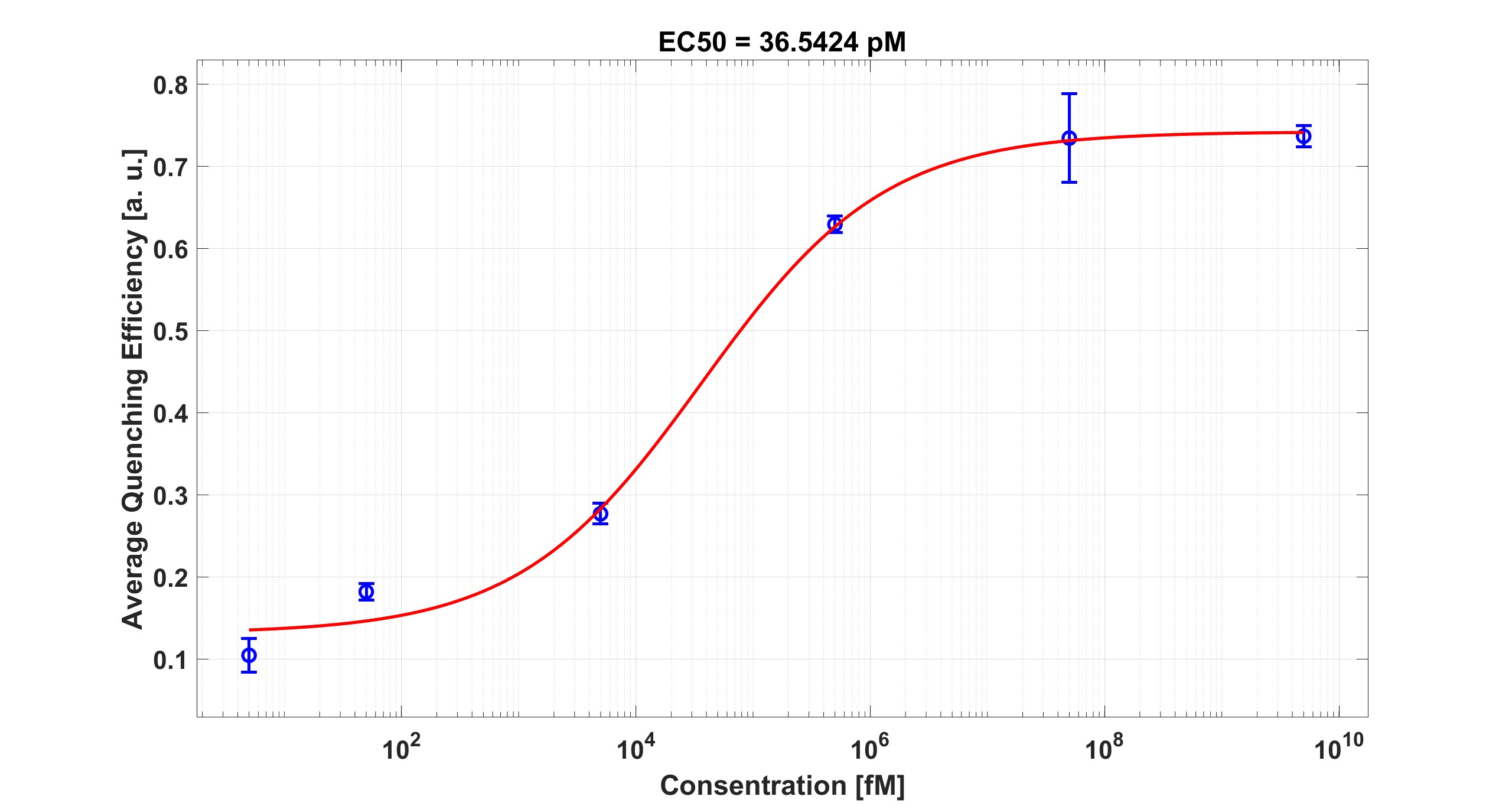}}
		\caption{Dose-response curve shows the quenching efficiency for each concentration tested, as shown in table 1.}\label{fig6}
	\end{figure}

To assess the dynamic range and limit of detection in this assay, 
UCNPs and AuNPs were mixed with \textit{SARS-CoV-2 cDNA} at different concentrations, as shown in table \ref{table1}. Figure \ref{fig6} shows the dose-response curve of this experiment. The lowest concentration tested was $5.06~fM$, and the highest concentration tested was $5.06~\mu M$. 
The quenching efficiency at $5.06~fM$ plus three times the standard deviation of the same data point was defined as the limit of the detection's quenching efficiency. 242 fM of the limit of detection was determined by fitting the dose-response curve to the 4-parameter logistic equation, and the EC50 point (midpoint between maximum and minimum quenching efficiency) was calculated to be 36.54 pM.
The red curve is the 4-parameter logistic curve:

\begin{equation}
    Y = Min + \frac{Max - Min}{1 + (\frac{x}{EC50})^{hc}}
    \label{equ1}
\end{equation}
where $Min$ and $Max$ are the minimum and maximum quenching efficiencies, $x$ is the concentration of DNA used in the measurement, $EC50$ is the concentration at which the quenching efficiency is $\frac{Max-Min}{2}$ and $hc$ is the hill coefficient. The $EC50$ of the fitted function was 36.54 pM 
and the hill coefficient of the fitted function was -0.5517 
The quenching efficiency for each concentration was defined as:
\begin{equation}
    QE = \frac{I_{neg}-I_{conc}}{I_{neg}}
    \label{equ1b}
\end{equation}
where $QE$ is quenching efficiency, $I_{neg}$ is the Intensity at 550 nm for the negative control test, $I_{conc}$ is the intensity at 550 nm for the tested concentration.

 One interesting question is how many UCNPs and AuNPs does it take to detect a single DNA molecule if the sample is positive? To determine the number of DNA molecules detected per UCNP for the positive sample, we need first to establish a detection limit, as illustrated in Fig. \ref{fig6}. The quenching efficiency increases from 10.4\% to 73.6\%, resulting in a contrast of 63.2\%. The midpoint of this efficiency range, 42\% quenching efficiency, corresponds to a DNA concentration of 36.54 pM, as determined by the four-parameter logistic curve fitted to our data. Assuming that a quenching efficiency above 42\% signifies a positive test, then the concentration of the DNA is known to be 36.54 pM. 
 
 This concentration, presented in the final reaction volume of 119 $\mu$L, is equivalent to $2.6\times10^{9}$ DNA molecules (supplementary materials). Given the constant number of UCNPs and AuNPs across all DNA concentrations, the positive test, as defined to have at least a concentration of 36.54 pM, reveals a specific DNA:UCNP:AuNP ratio for detecting a single DNA to be, calculated to be $1:1.8\times10^{3}:9\times10^{3}$. Therefore, the detection of a single DNA molecule for a positive test as defined above necessitates a $1.8\times10^{3}$ UCNPs and $9.0\times10^{3}$ AuNPs. Note that we kept the ratio of UCNPs to AuNPs constant and equal to 1:5 in the entire experiment.

To test the specificity of our assay, we prepared a different mismatched DNA sequence. By design, half of the \textit{SARS-CoV-2 cDNA} hybridizes with UCNP's primer (21 bases), and the remaining half hybridizes with the AuNP's primer (remaining 21 bases). In this case, we prepared a mismatch DNA in which 12 bases are mismatched on each side (DNA-mmP1P2, containing 24 mismatch bases out of 42), a mismatch DNA in which only 12 mismatch bases are on the UCNP side while there are no mismatch bases on the AuNp side (DNA-mmP2, containing 12 mismatch bases out of 42), and a mismatch DNA in which only 12 mismatch bases are on the AuNp side while there are no mismatch bases on the UCNP side (DNA-mmP1, containing 12 mismatch bases out of 42) (see supplementary materials for sequences). When the mismatched bases are present on both sides, as in DNA-mmP1P2, the quenching of the 550 nm peak results in a reduction of 2984 counts (39.3 \% quenching efficiency). 

\begin{figure}[H]
     \centering
     \begin{subfigure}[b]{1\textwidth}
         \centering
         \includegraphics[width=.95\textwidth]{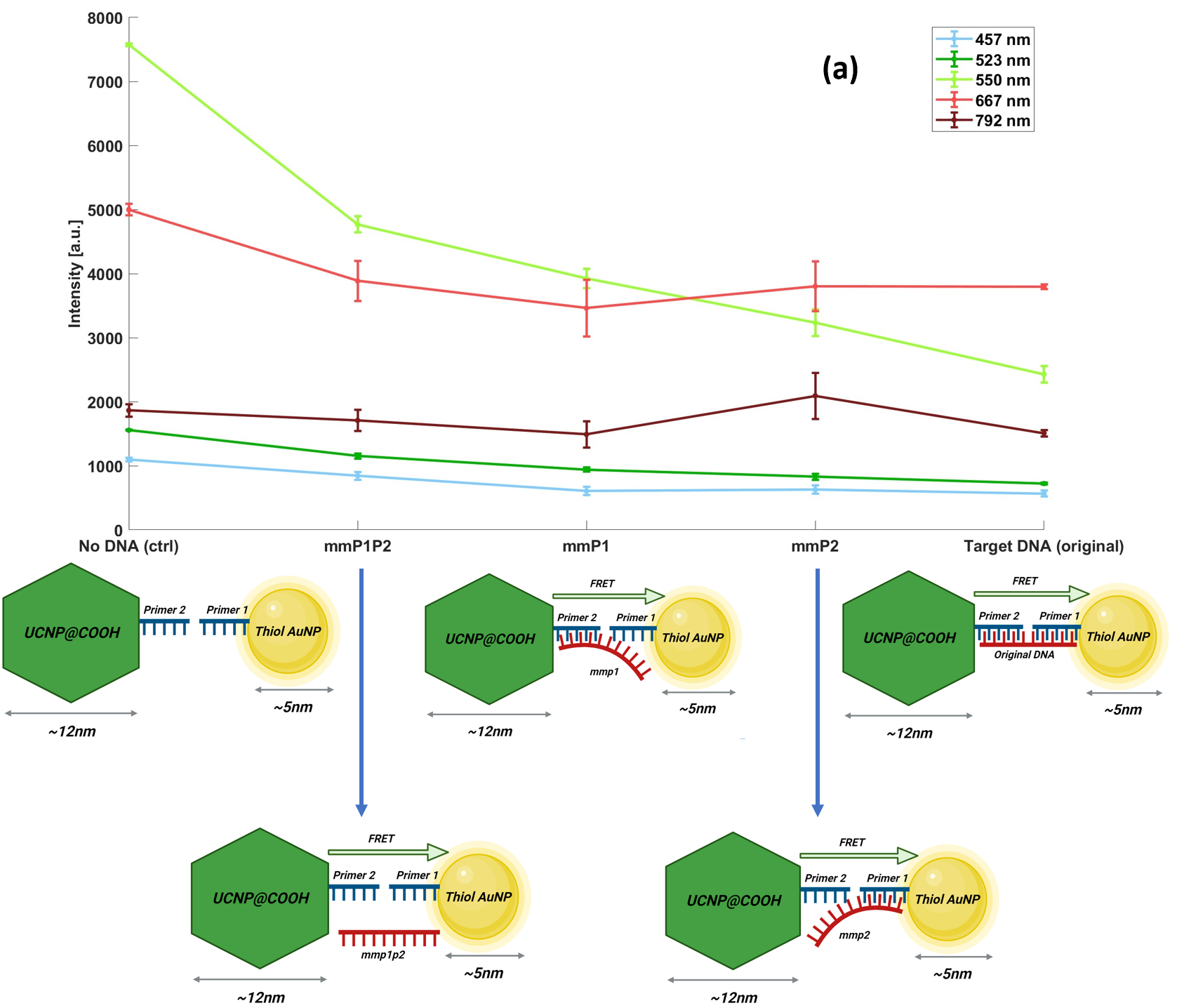}
         \label{fig7-A}
     \end{subfigure}
     \begin{subfigure}[b]{1\textwidth}
         \centering
         \includegraphics[width=.75\textwidth]{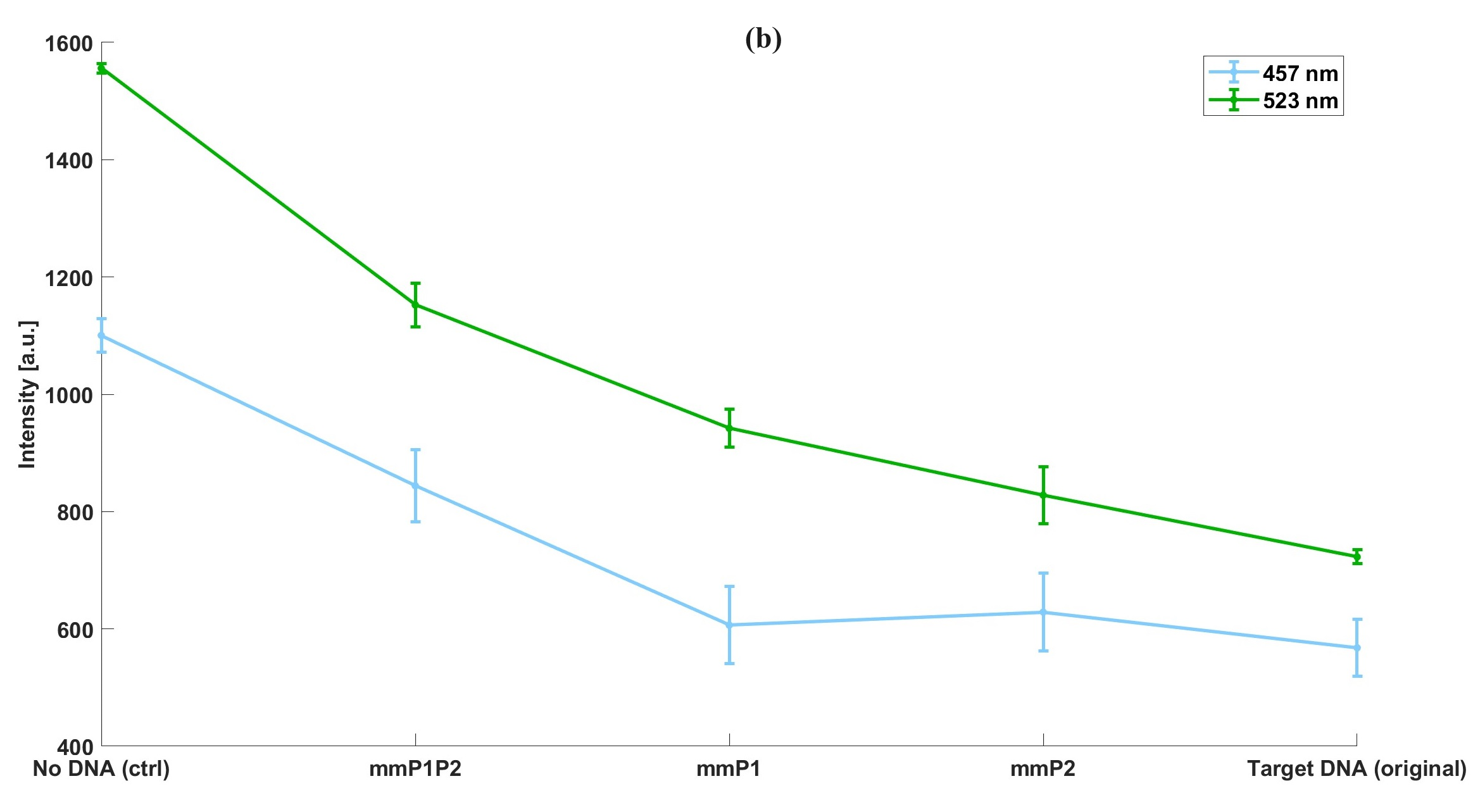}
         \label{fig7-B}
     \end{subfigure}
        \caption{Results of the miss-matched DNA experiment for determination of specificity. A) shows the average intensity at major UCNP peaks for each type of miss-match DNA sequence. The target DNA shows the maximum quenching compared to the negative control, measured at 64.4 \%, while the mmP1P2, where the 12 miss-match bases are on the UCNP side, and 12 miss-match bases are on the AuNp side as well, shows only 39.3 \%. (b) shows the zoomed-in view for 523 nm and 457 nm results, both showing much lower sensitivity compared to 550 nm.}
        \label{fig7}
\end{figure}

In contrast, the original target DNA, which has no mismatches, exhibits a more pronounced decrease of approximately 4907 counts in intensity (64.6 \% quenching efficiency), both values being relative to the negative control (see Fig. \ref{fig7}). Meanwhile, in the case of DNA-mmP1, where the DNA is only supposed to hybridize with primer 2 on the UCNP side, the intensity of 550nm wavelength is observed to be 3785 counts, a 3806 reduction relative to the negative control (quenching efficiency of 50.1\%). Furthermore, in the case of DNA-mmP2, where the DNA is only supposed to hybridize with primer 1 on AuNPs, the intensity of 550 nm wavelength is observed to be 3498 counts, a reduction of 4093 counts (quenching efficiency of 53.9\%).

Several factors can cause such differences. First, the affinity of each mismatched DNA sequence is different from the two primers. This could cause different levels of quenching for mismatched tests. The second reason could be the high concentration of the DNA used. Even for relatively lower affinities (in terms of the dissociation constant $K_{D}$), higher concentrations can still lead to a high binding ratio (ratio of bound DNA to total DNA). Further studies are needed to understand the dependency of the quenching efficiency to the mismatch sequence and the location of mismatch bases. Furthermore, the quenching behavior of 457 nm and 523 nm wavelengths are similar to 550 nm, while the 792 nm and 667nm peaks do not show any significant quenching for any of the samples. The theoretical binding affinity of mismatch DNA sequences and the primers are in table S2. The target DNA has the highest affinity (lowest Kd Value) indicated by the Kd value of $~10^{-42}\ M$. Next is the mmP1 with Kd value of $~10^{-31}\ M$, then mmP2 with Kd value of $~10^{-27}\ M$. Lastly, it is mmP12 with Kd value of $~10^{-16}\ M$. 

An important indication of this test is that the specificity of such a test depends on the exact location of the mismatched bases. This can be inferred from the difference between DNA-mmP2 and DNA-mmP1P2 or DNA-mmP2 and DNA-mmP1. It can be seen that the mismatch on the UCNP side has less effect on the hybridization compared to the other two mismatched samples.

To interpret the results of the mismatch experiments, it's essential to understand that quenching arises from the binding of gold nanoparticles (AuNPs) with upconversion nanoparticles (UCNPs). This binding depends on the affinity of the DNA to be detected and its concentration relative to the nanoparticle-bound primers, which are at a constant concentration. We can apply the Langmuir equation to accurately quantify DNA binding to the primers without compromising generality, accuracy, or precision:
\begin{equation}
\theta = \frac{[\text{DNA.Primer}]}{[\text{Primer}]_{\text{total}}} \approx \frac{[\text{DNA}]_{\text{total}}}{K_d + [\text{DNA}]_{\text{total}}}
\end{equation}
where \([\text{DNA.Primer}]\) represents the concentration of DNA bound to both primers, \([\text{Primer}]_{\text{total}}\) denotes the total concentration of primers (i.e., UCNPs or AuNPs, with appropriate unit conversion), and \([\text{DNA}]_{\text{total}}\) is the total DNA concentration (target, mmP1, mmP2, or mmP1P2). The critical question is: what quenching efficiency should we anticipate for the mmP1, mmP2, and mmP1P2 mismatched DNAs compared to the target DNA and among each other? Specifically, how does the quenching efficiency order from highest to lowest?

To address this question, given that mismatched bases alter the dissociation constant \( K_d \), we calculated \( K_d \) values for each mismatch sequence and the target DNA for comparison. An online tool is available at (http://biotools.nubic.northwestern.edu/OligoCalc.html)\citep{oligocalc} was utilized for these calculations, with results detailed in the supplementary materials. The relationship between changes in free energy (\( \Delta G \)) and \( K_d \) is expressed as:
\begin{equation}
K_d = \frac{[\text{DNA}][\text{Primer}]}{[\text{DNA.Primer}]}
\end{equation}
\begin{equation}
\Delta G = RT \ln\left(\frac{[\text{DNA.Primer}]}{[\text{DNA}][\text{Primer}]}\right) = RT \ln\left(\frac{1}{K_d}\right)
\end{equation}
We inputted sequences excluding the mismatch bases for our DNA samples, enabling the application to calculate changes in free energy as if the DNA were to bind with its exact hybrid. In reality, however, mismatch bases persist and are hypothesized to decrease, rather than increase, affinity. The resulting \( K_d \) values are presented in the table below.

\begin{table}[H]
\centering
\caption{Calculated changes in free energy and binding affinities for the original target DNA and the mismatches.}
\label{your-table-label}
\begin{tabular}{|l|c|c|c|}
\hline
DNA     & $\Delta G$ (Kcal/mol)  & $\Delta G/RT$ & $K_d$ (exp($-\Delta G/RT$)) [M]\\ \hline
Target  & 56.8                   & 95.4          & $3.7 \times 10^{-42}$          \\ 
mmP1    & 41.5                   & 69.7          & $5.4 \times 10^{-31}$          \\ 
mmP2    & 36.8                   & 61.8          & $1.4 \times 10^{-27}$          \\ 
mmP1P2  & 21.5                   & 36.1          & $2.1 \times 10^{-16}$          \\ \hline
\end{tabular}
\end{table}

The affinity order of the DNA samples is evident, with lower \( K_d \) values indicating higher affinity: Target DNA > DNAmmP1 > DNAmmP2 > DNAmmP1P2. Consequently, we would anticipate the quenching strength of the samples, from weakest to strongest, to be DNA-mmP1P2, DNA-mmP2, DNA-mmP1, and, ultimately, target DNA. However, it's important to note that our calculations of DNA affinity exclude mismatch bases, leading to an inherent overestimation of affinity in the results. Remarkably, the \( K_d \) values are exceedingly small, with the highest being \( 2.1 \times 10^{-16} \) M for DNA-mmP1P2. Given our experiment's mismatch binding concentration of \( 5.06\times 10^{-6} \) M, complete quenching suppression is unlikely. This implies that despite a relatively higher \( K_d \) value and assumed lower primer affinity, significant binding activity is still observed, albeit less than that of target DNA, as shown in Fig. \ref{fig7}.

In this experiment, we are detecting the number of UCNP-primer 2-mismatch DNA-primer 1-AuNP complexes by measuring the spectrum of the complex and comparing it with negative control via equation \ref{equ1b}. The major issue is that the more complexes we have, the higher the quenching efficiency will be. But the number of these complexes in a given sample depends on three factors: 1) the concentration of mismatch DNA, 2) the affinity of mismatch DNA with primer 1, 3) the affinity of mismatch DNA with primer 2. In our mismatch measurements Fig. \ref{fig7}, we kept all DNA concentrations constant and at $5.06 \mu M$ and expected the change in quenching efficiency to be only due to the affinity of DNA and primers. A keen reader may argue that the mismatch DNA mmP1P2, where the DNA has 12 mismatch bases on each primer side, should result in 0\% quenching efficiency, or in terms of Fig. \ref{fig7} similar intensity as negative control. However, our results do not show that. This is because, no matter how low the affinity of DNA and primers may be assumed, there will be a concentration that the ratio of DNA-primer complexes to total available primers will be 1.

The Langmuir equation provides insight into this argument:
\[ \theta = \frac{[\text{DNA.Primer}]}{[\text{Primer}]} \approx \frac{[\text{DNA}]_{\text{total}}}{K_d + [\text{DNA}]_{\text{total}} } \]
Considering the substantial difference between \( K_d \approx 10^{-16} \) M and a concentration of \( 5.06\times10^{-6} \) M, the theoretical value of \( \theta \) approaches 1. However, it's important to acknowledge that this theoretical model may not fully align with real-world observations due to various factors like temperature, pH, measurement errors, human error, and so on. Despite this, the model sets a certain expectation and explains why we observed quenching with DNA-mmP1P2, although less than with target DNA.

\begin{figure}[H]
	\centerline{\includegraphics[width=1.1\textwidth,clip=]{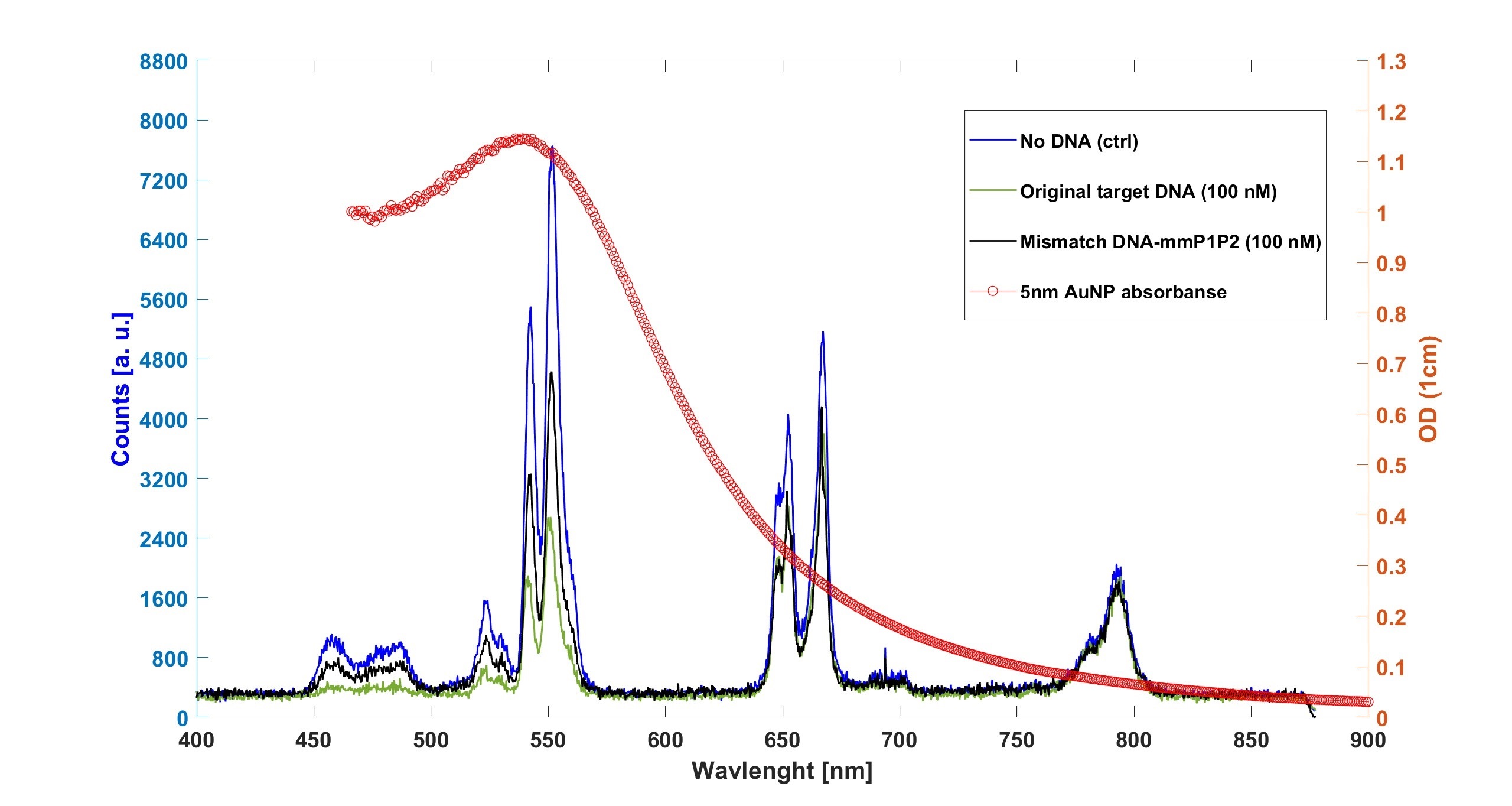}}
	\caption{Spectrum of the negative control test, positive control test, and DNA mmP1P2.}\label{fig8}
	\end{figure}

The mismatch experiment revealed that alterations in the sequence at both hybridization sites of the selected SARS-CoV-2 cDNA, corresponding to primer 1 and primer 2, lead to a significant reduction in quenching efficiency, approximately by 40 percent, as illustrated in Fig. \ref{fig8}. However, mutations occurring on only one side of the sequence result in a less pronounced decrease in efficiency. Specifically, when the mutation is located on the side where the primer for the AuNPs binds, the test maintains its effectiveness relatively well.

\begin{figure}[H]
	\centerline{\includegraphics[width=1.1\textwidth,clip=]{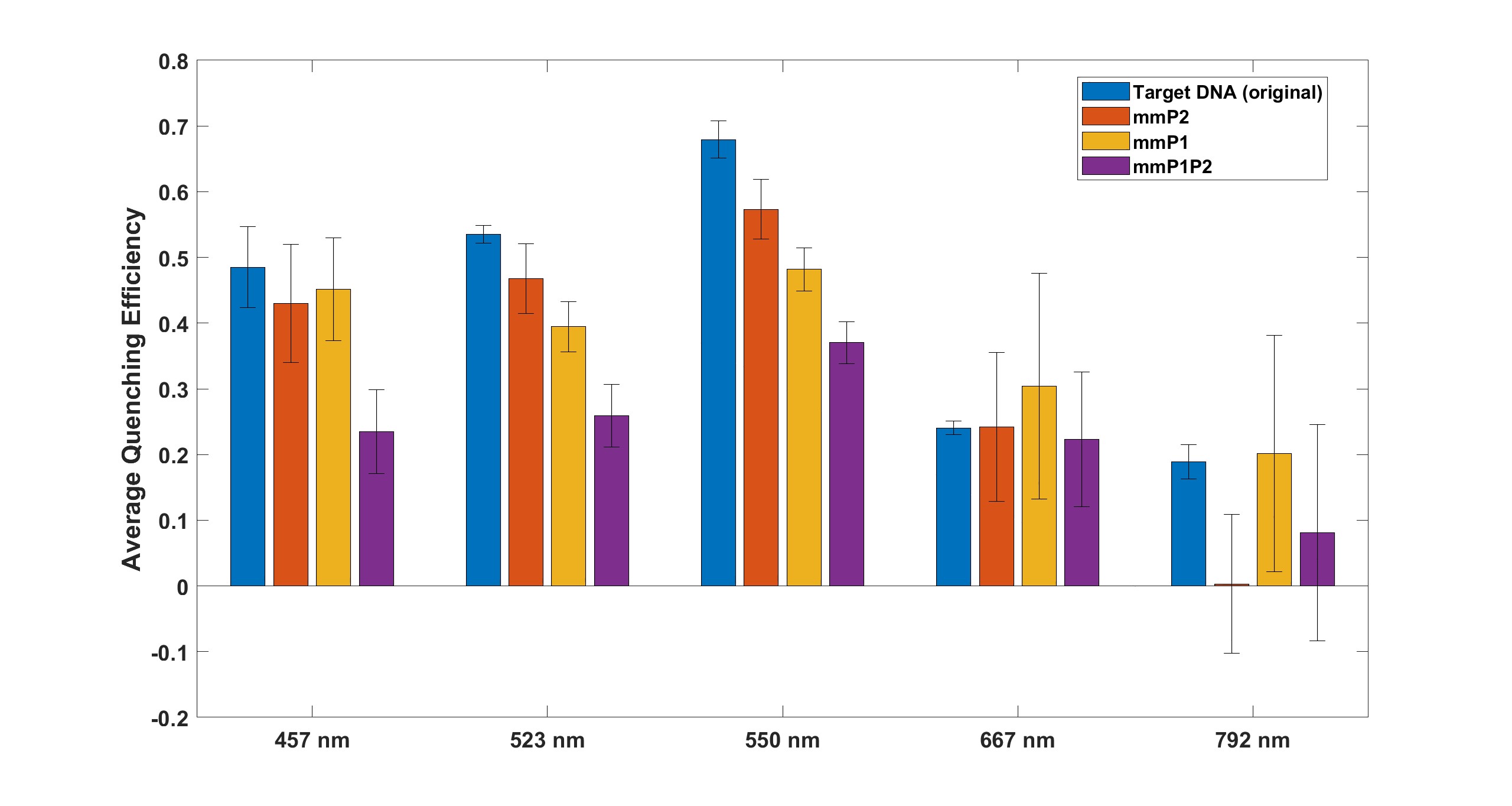}}
	\caption{Comparative Analysis of Quenching Efficiencies across Different Wavelength Bands.}\label{fig9}
	\end{figure}

  We also examine the quenching efficiencies across various bands in our experiment for both the original and mismatched DNA. The quenching efficacy across all color bands has been charted for a comprehensive understanding, as illustrated in the accompanying figure. Notably, 792 nm does not exhibit substantial quenching in contrast to other bands. This observation is in alignment with the AuNP absorbance, where greater quenching is observed in the green and blue bands, with less quenching evident in the red band and minimal quenching at 792 nm. These findings bolster the evidence for the occurrence of Luminescence Resonance Energy Transfer (LRET) in our experiment. Figure \ref{fig9} provides a visual representation of these findings. 

\section{conclusions}
In this study, we have demonstrated an efficient and highly specific assay for DNA detection, showcasing its utility in identifying SARS-CoV-2 cDNA. Utilizing upconversion nanoparticles (UCNPs) conjugated with one primer and gold nanoparticles (AuNPs) conjugated with another, we engineered a system where these nanoparticles hybridize with opposite ends of the SARS-CoV-2 cDNA. This design led to the binding of multiple AuNPs to each UCNP, which in turn varied the quenching of UCNP fluorescence. Notably, the upconverted fluorescence of our UCNPs, emitting wavelengths of 426, 524, 551, 666, and 792 nm, quenched following the AuNP's absorption spectrum, with peak quenching observed at 551 nm and minimal quenching at 792 nm. This quenching mechanism is attributed to luminescence resonance energy transfer (LRET) coupling, facilitating energy transfer from the UCNPs to the AuNPs.

Our assay successfully detected SARS-CoV-2 cDNA at concentrations as low as 242 fM. The assay's specificity was further confirmed through testing with various mismatched DNA sequences, revealing a decrease in quenching efficiency upon the introduction of mismatch bases on both the DNA's UCNP and AuNP binding sides. This result indicates the assay's sensitivity to the precise DNA sequence, an essential feature for accurate viral detection. The midpoint of quenching efficiency at 42\% 
corresponding to a DNA concentration of 36.54 pM, defined positive test threshold concentration, underscores the assay’s potential for highly sensitive applications.

\section*{Acknowledgements}
Multiple esteemed organizations supported this research, and we extend our sincere gratitude to each for their invaluable contributions. We acknowledge the financial support from the Air Force Office of Scientific Research (Award No. FA9550-20-1-0366), Office of Naval Research (Award No. N00014-20-1-2184), and the Robert A. Welch Foundation (Grants No. A-1261, A-1547). Our work also benefited from grants from the National Science Foundation (Grant No. PHY-2013771, PHY-1820930, ECCS-2032589), and the National Institutes of Health (Award No. R03AI139650 and R21AI149383).

This material is based on work supported by the U.S. Department of Energy, Office of Science, and Biological and Environmental Research under Award Number DE-SC-0023103. Special acknowledgment is given to the Hagler Institute for Advanced Study at Texas A\&M University for their support of SE's involvement in this research. Additionally, we thank the Herman F. Heep and Minnie Belle Heep Texas A\&M University Endowed Fund, held and administered by the Texas A\&M Foundation, which supported AH's contributions. Each organization's dedication to promoting scientific research has been crucial to our endeavors, and we are profoundly thankful for their support.


	\bibliographystyle{abbrvnat}
	\setcitestyle{authoryear}
	\bibliography{library}
\end{document}